\documentclass[aps, prl, reprint, showkeys, nolongbibliography]{revtex4-2}

\pdfoutput=1

\usepackage[utf8]{inputenc}
\usepackage[usenames,dvipsnames]{color}
\usepackage{hyperref}
\usepackage{amssymb}
\usepackage{amsmath}
\usepackage{multirow}
\usepackage{epsfig}
\usepackage{rawfonts}
\usepackage{latexsym}
\usepackage{theorem}
\usepackage{natbib}
\usepackage{graphicx}

\begin{document}

\title{Quantum Monte Carlo Study of Semiconductor Artificial Graphene Nanostructures}
\date{\today}

\author{G\"{o}khan \"{O}ztarhan}
\affiliation{Department of Physics, \.{I}zmir Institute of Technology, 35430 Urla, \.{I}zmir, Turkey}
\author{E. Bulut Kul}
\affiliation{Department of Physics, \.{I}zmir Institute of Technology, 35430 Urla, \.{I}zmir, Turkey}
\author{Emre Okcu}
\affiliation{Department of Physics, \.{I}zmir Institute of Technology, 35430 Urla, \.{I}zmir, Turkey}
\author{A. D. G\"{u}\c{c}l\"{u}}
\affiliation{Department of Physics, \.{I}zmir Institute of Technology, 35430 Urla, \.{I}zmir, Turkey}

\begin{abstract}
  Semiconductor artificial graphene nanostructures where Hubbard model
  parameter $U/t$ can be of the order of 100, provide a highly
  controllable platform to study strongly correlated quantum
  many-particle phases. We use accurate variational and diffusion
  Monte Carlo methods to demonstrate a transition from
  antiferromagnetic to metallic phases for experimentally accessible
  lattice constant $a=50$ nm in terms of lattice site radius $\rho$,
  for finite sized artificial honeycomb structures nanopatterned on
  GaAs quantum wells containing up to 114 electrons. By analysing
  spin-spin correlation functions for hexagonal flakes with armchair edges
  and triangular flakes with zigzag edges, we show that edge type, geometry
  and charge nonuniformity affect the
  steepness and the crossover $\rho$ value of the phase
  transition. For triangular structures, the metal-insulator
  transition is accompanied with a smoother edge polarization
  transition.
\end{abstract}


\maketitle

In recent years, technological advances in photonic and condensed
matter based artificial superlattices gives us opportunities to
develop practical quantum
simulators \cite{Bloch2005,Mazurenko2017,Buluta2009,10.1093/nsr/nww023,Weimer2010,Islam2011,Salfi2016,PhysRevX.8.031022,Aspuru-Guzik2012,Cai2013,Bernien2017,li21,AG_low_disorder,AG_observation_of_dirac_bands}.
These quantum simulators allow us to replicate complex systems that
are hard to fabricate and provide us with a playground to verify
theoretical predictions.  In this respect, artificial graphene (AG)
nanostructures, designed by imitating the 2D honeycomb pattern of
graphene, have been proven to be good candidates for being reliable
and controllable sources for both fabrication and investigation of
many physical phenomena related to Dirac fermions
\cite{Gomes2012,Tarruell2012,Lu2014,Jackmin2014,AG_low_disorder,AG_observation_of_dirac_bands,Wang2016_AG}.
In particular, AG nanostructures can be formed using semiconductor
materials. While earlier reports on nanopatterned artificial graphene
on GaAs quantum well (QW) structures found no evidence of massless
Dirac fermions (MDFs), presumably because of the relatively large
lattice periods \cite{N_dvorn_k_2012,AG_theoretical_0,Singha2011}, in
recent experimental works using modulation-doped AlGaAs/GaAs quantum
wells \cite{AG_observation_of_dirac_bands,AG_low_disorder} shrinking
down of the lattice constant of the honeycomb array to approximately
50 nm allowed the observation of the predicted graphene-like
behavior \cite{AG_theoretical_0,AG_theoretical_1,AG_theoretical_2,AG_theoretical_3}.

Observation of Dirac fermions in AG also opens up a fresh way of
studying graphene quantum dots \cite{graphene_devrim_book} where
geometry, size and edge type is expected to give rise to several
physical properties such as bandgap opening \cite{GQD_devrim_bandgap},
edge magnetization \cite{Rossier2007,Kaxiras2008,GQD_devrim_triangle}
and optical control \cite{GQD_devrim_optics}. However, the fabrication
and reliability issues such as edge reconstruction or presence of
impurities, make it harder to observe interesting phenomena predicted
to occur in nanostructured graphene. Semiconductor AG nanostructures,
on the other hand, offer several advantages such as tunability of
system parameters including lattice constant, site radius and
potential depth, which, in turn, allow to control electron-electron
interactions and tunneling strength between sites, in particular
Hubbard parameter $U/t$. In experimental structures with lattice
constant $a=50$ nm \cite{AG_observation_of_dirac_bands}, $U/t$ can be
as high as 350 (as we will argue below) i.e. two orders of magnitude
larger than the critical value for antiferromagnetic
Mott transition predicted by calculations based on Hubbard
model for honeycomb lattice
\cite{hubbard_honeycomb_0,hubbard_honeycomb_1,hubbard_honeycomb_2,hubbard_honeycomb_3}.
Moreover, unlike in real graphene, long
range electron repulsion does not cancel the attraction of the
artificial confining potential in AG even near charge
neutrality. For such large and long-ranged electron interactions, a
non-perturbative many-body approach is desirable for careful treatment
of correlation effects.

Earlier theoretical work on electron interaction effects in
semiconductor AG nanostructures based on density functional theory
(DFT) showed that Dirac bands were stable against
interactions \cite{AG_theoretical_1,AG_theoretical_2} which was also
confirmed using path integral Monte Carlo
calculations \cite{AG_theoretical_3}. However, recent calculations
using Hartree-Fock and exact diagonalization approaches for a
triangular zigzag geometry with $a=12.5-15$ nm show that a transition
from antiferromagnetic (AF) insulator to metallic phases occurs, pointing to
the importance of electron interactions \cite{AG_theoretical_yasser}.

In this work, we use continuum variational Monte Carlo (VMC) and diffusion
Monte Carlo (DMC) methods for non-perturbative and accurate treatment of many-body
correlations within the fixed-node approximation to study GaAs based
AG nanostructures. First, we consider an hexagonal armchair geometry
which serves as a bridge between the finite-size samples and bulk
graphene \cite{GQD_devrim_bandgap}, with lattice constant $a=50$ nm
following recent experimental
work \cite{AG_observation_of_dirac_bands}. We show that, a transition
from AF to metallic phase occurs, but is affected by a nonuniform
charge distribution in the sample due to finite size effects.  This
charge nonuniformity which is not present in real graphene quantum
dots, causes the phase transition to be steeper and to occur at a
smaller value of $\rho$. We also investigate AG quantum dots with
triangular zigzag geometry and show that edge magnetization survives
the phase transition, in agreement with previous theoretical
prediction for smaller lattice constants \cite{AG_theoretical_yasser}.

Our model of nanostructured semiconductor AG consists of $N$
interacting electrons in a honeycomb array of $N$ confining
potentials, described by the many-body Hamiltonian
\begin{align} \label{hamiltonian}
  H = -\frac{1}{2} \sum_{i}^N \nabla^{2}_{i} + \sum_{i}^N V(\textbf{r}_{i})
      + \sum_{i}^N k |\textbf{r}_{i}|^2 + \sum_{i<j}^N \frac{1}{r_{ij}}
\end{align}
in effective atomic units (electronic charge $e$, dielectric constant $\epsilon$,
effective mass $m^*$, and $\hbar$ are set to 1),
where $1 / r_{ij}$ is the Coulomb interaction between the electrons,
$V(\textbf{r}_{i})$ is the total potential energy of the confining
potentials, and $k$ is the spring constant of quadratic gate potential
located at the center of the system which controls the finite size
effects. Typical material properties for GaAs, effective electron mass
$m^{*} = 0.067 m_{0}$ and dielectric constant $\epsilon = 12.4$, are
used.  Corresponding effective Bohr radius is $a^{*}_{0} = 9.794$ nm,
and the effective Hartree energy is $11.857$ meV. The honeycomb array
of potential wells is modelled using gaussian-like
functions \cite{AG_theoretical_3},
\begin{align} \label{potential}
    V(\textbf{r}) = V_{0} \sum\limits_{\textbf{R}_{0}}
    \exp[-(|\textbf{r}-\textbf{R}_{0}|^{2} / \rho^{2})^{s}]
\end{align}
where $s \geq 1$, $V_{0}$ is the potential depth, $\rho$ is the radius
and $s$ is the sharpness of the potential wells. $\textbf{R}_{0}$ is
the location of the potential wells. In our numerical calculations,
dot-to-dot distance (lattice constant) was fixed to $a=50$ nm, while
several radius values from $\rho=10$ nm to $35$ nm were covered. Three
different sharpness values were used; $s = 1$ for a gaussian
potential, $s = 2.8$ for a sharp, muffin-tin like potential, and
$s = 1.4$ in between. $V_0$ values vary depending on dot radius
(e.g. increasing monotonically with dot radius from $-38$ meV
to $-15$ meV for $s = 1.4$ and $N=42$),
tuned to keep the total energy of the system
close to zero, since our aim is to imitate the charge neutral behavior
of finite sized graphene quantum dot. For too high values of $V_0$,
electrons tend to escape the system during VMC or DMC simulations,
while for too low values over localization
occurs \cite{AG_theoretical_0}.

Accuracy of the numerical calculations depends on trial wave functions
in both VMC and DMC methods. One starts with a set of single-particle
orbitals (e.g. localized gaussians or from self-consistent
calculations) to build a Slater-Jastrow trial wave function
$\Psi_{T}(\textbf{R})$ which is a linear combination of products of
up- and down-spin Slater determinants of these orbitals multiplied by
a Jastrow factor (The details of our Jastrow factor is given in
Ref. \onlinecite{qmc_devrim_0}). After the VMC calculations where
Jastrow parameters as well as the gaussian functions width are
optimized using energy minimization technique \cite{qmc_cyrus_2}, we
use fixed-node DMC \cite{qmc_review,qmc_cyrus_1} to project the
optimized many-body wave function onto a better approximation of the
true ground state, an approximation that has the same nodes as
$\Psi_{T}(\textbf{R})$. The resulting fixed-node DMC energy is an
upper bound to the true energy and depends only on the nodal structure
of the Slater part of the trial wave
function $\Psi_{T}(\textbf{R})$ \cite{qmc_review}.

In order to form Slater determinants, we prepare three different types
of orbitals aiming to capture metallic or AF insulator
phases, depending on the potential well radius $\rho$: (i) Localized
gaussian functions that are proven to be one of the most suitable
functions for 2D systems of quantum dots \cite{qmc_devrim_0,
  qmc_devrim_1, qmc_devrim_2} and are expected to provide a better
description of strongly localized states. (ii) Tight-binding (TB) orbitals,
on the other extreme, may be used to describe metallic phases in which
electrons move more freely. (iii) Mean-field Hubbard (MFH) orbitals can
describe both localized and liquid-like states depending on the ratio
$U/t$. Corresponding variational and fixed-node energies of those
three types of orbitals are expected to hint us at a possible
transition from metallic state to an AF order as a
function of $\rho$.  In this work, all quantities that do not commute
with the Hamiltonian were calculated using an extrapolated estimator,
$\langle \hat{O} \rangle = 2\langle \hat{O} \rangle_{DMC} - \langle
\hat{O} \rangle_{VMC}$ \cite{qmc_review}.

\begin{figure}[h]
\centering
\includegraphics[width=0.950\columnwidth]{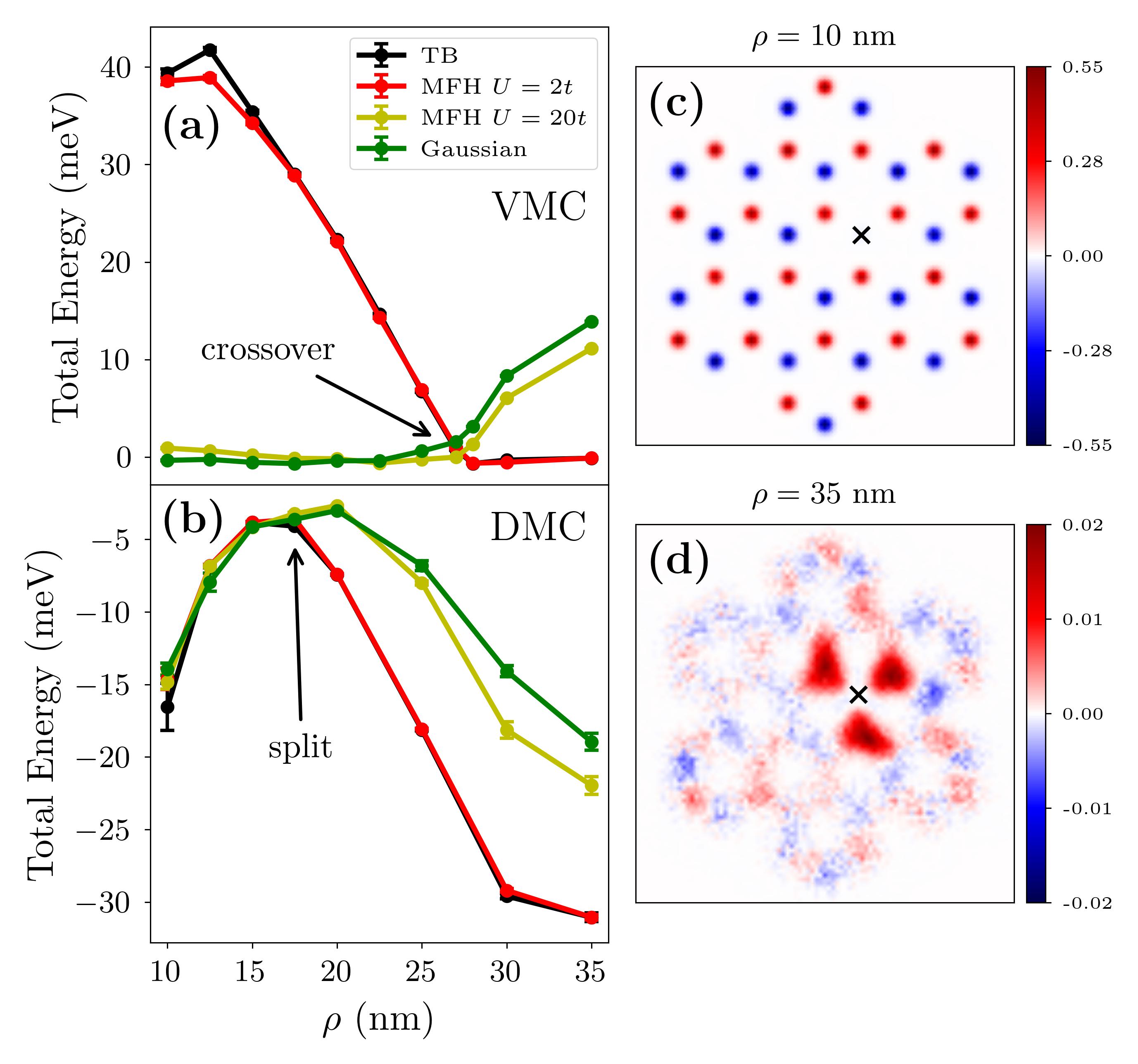}
\caption{\label{first_result} 
    Armchair hexagonal flake results,
    $s=1.4$, $k=0$ and $N=42$ plotted for several trial wave functions.
    (a) VMC total energy vs $\rho$.
	(b) DMC total energy vs $\rho$.
    (c) Extrapolated pair spin density results for $\rho = 10$ nm, using tight-binding trial wave function.
    (d) Extrapolated pair spin density results for $\rho = 35$ nm, using tight-binding trial wave function.
    The reference electron is located at the X marked point for c and d.
}
\end{figure}
Figure \ref{first_result}a shows the VMC energies obtained from TB,
MFH and gaussian orbitals, as a function of $\rho$, for
an armchair hexagonal geometry with 42 sites and 42 electrons.  The
parabolic gate potential parameter is turned off ($k = 0$) and its
effect will be discussed in a later section. At low $\rho$, gaussian
and MFH $U=20t$ orbitals provides better variational energies. As
$\rho$ is increased to $\approx 27$ nm, a clear crossover (from
insulator to metallic phase) occurs above which TB and MFH $U=2t$
orbitals take the lead. Fixed-node DMC energies, however, reveal a
somewhat different picture, shown in Fig.\ \ref{first_result}b.  From
$\rho=10$ nm to $\rho \approx 18$ nm, all trial wave functions give
similar energies within the statistical error bars, and split near
$\rho \approx 18$ nm. After the split, the ground state of the system
is represented by TB and MFH $U=2t$ trial wave functions. These
results show that, surprisingly, TB trial wave function has equally
good nodal structure as the gaussian orbitals at low $\rho$ values,
raising questions about the true nature of the ground state. To reveal
the underlying electronic and magnetic structure, we consider the pair
densities $p_{\sigma \sigma_0}( \textbf{r},\textbf{r}_0)$, probability
of finding an electron with spin $\sigma$ at location $\textbf{r}$
when an electron with spin $\sigma_o$ is fixed at location
$\textbf{r}_0$, and the pair spin densities,
$p_{\uparrow\downarrow }( \textbf{r},\textbf{r}_0)-p_{\downarrow
  \downarrow }( \textbf{r},\textbf{r}_0)$. Figures
\ref{first_result}c-d show the pair spin densities for a reference
spin down electron fixed on top of a site (chosen to break the system
symmetry and away from the edges) shown with a cross. At $\rho=10$ nm,
electrons are well localized at the sites, leading to an
AF insulator. On the other hand, at large
values of $\rho$, spin-spin correlations are weak and short-ranged up
to nearest neighbors.  While these results confirm that a transition
from AF insulator to a metallic phase does occur as
observed in previous work \cite{AG_theoretical_yasser}, there are
several issues left to address: (i) Inconsistent signature regarding
the crossover $\rho$ value obtained from VMC and DMC energies and the
underlying dynamics of the transition. (ii) Finite size and edge
effects. (iii) Relationship to bulk properties.

\begin{figure}[h]
\centering
\includegraphics[width=0.925\columnwidth]{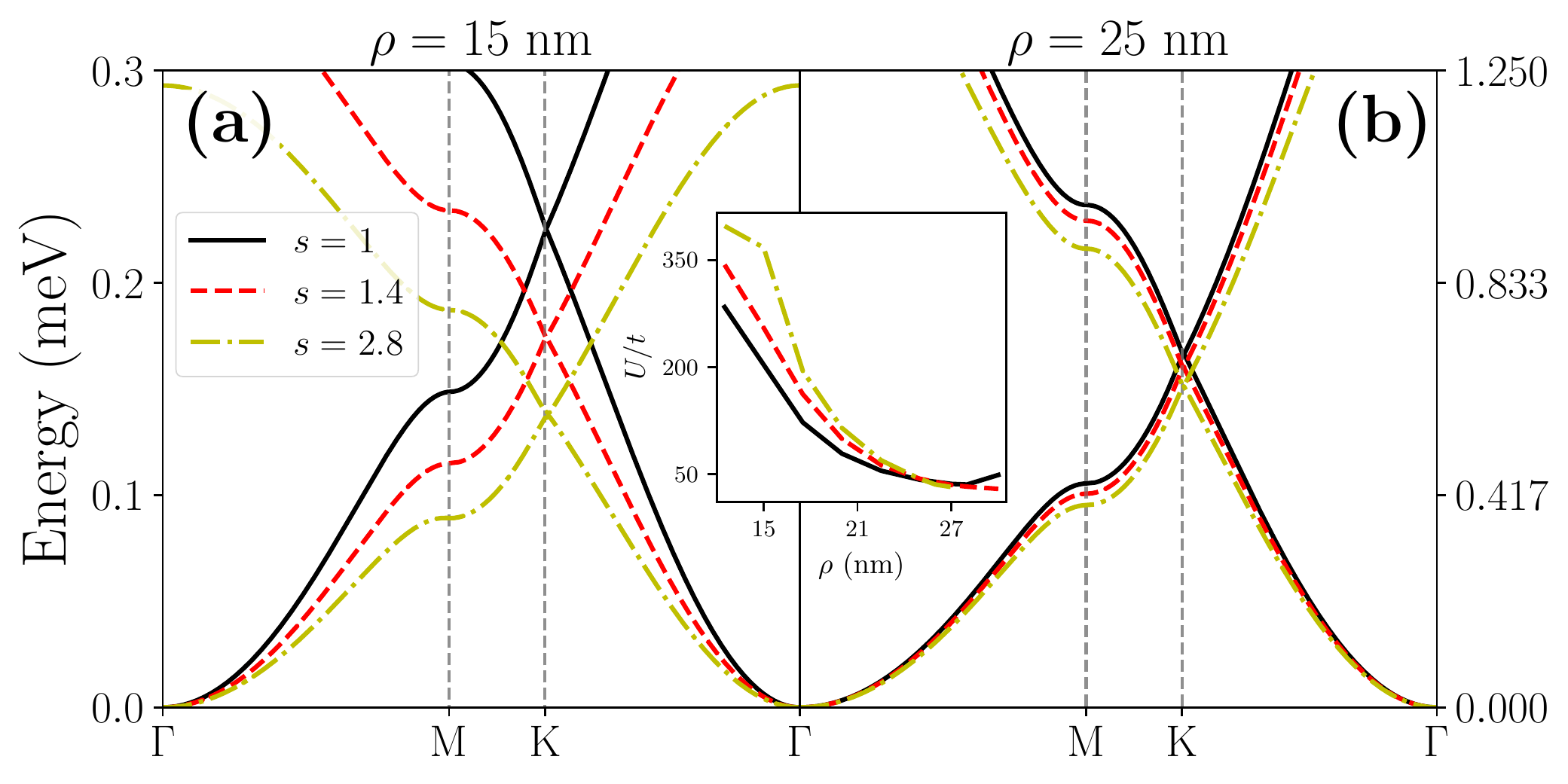}
\caption{\label{dft} 
    DFT band structure of bulk artificial graphene for (a) $\rho = 15$ nm and (b) $\rho = 25$ nm.
    Inset figure shows that $U/t$ ratio plotted against dot radius $\rho$, predicted by DFT and single particle calculations.
}
\end{figure}
Now, we turn our attention to the 2D bulk properties for similar
system parameters used in our calculations. As mentioned earlier,
$V_0$ values are tuned to keep the total energy of the system close to
zero. The question is whether such set of parameters leave the Dirac
cone structure intact, which may otherwise have repercussions on
electronic and magnetic properties. In Fig.\ \ref{dft}, we show our
DFT calculations results in the local density approximation (LDA). The
Dirac cone structure is preserved for the range $\rho = [12.5, 30]$ nm;
outside this range the Dirac fermion picture becomes distorted.
Moreover, we can estimate the TB hopping
parameter $t$ from the slope of the Dirac cones as
$t = \frac{2}{3} a^{-1} dE / dk$ where $a$ is the lattice constant.
As we see from Fig.\ \ref{dft}, $t$
increases with increasing $\rho$ but decreasing $s$, as expected. In
addition to $t$, to build a Hubbard model, we estimated the $U$ value
for a single well using $U = 2 \pi \int r n(r) V_{e}(r) dr$ after
solving single particle Schr\"{o}dinger equation. In the inset of
Fig.\ \ref{dft}, we plot $U/t$ as a function of $\rho$ which shows a
fast decay from $\approx 300$ to 50 between $\rho=15$ and $20$.
Surprisingly, these results indicate that the
critical $U/t$ value for metal-insulator transition in AG is much
higher than the critical value of $\sim 3.8$ predicted by Hubbard calculations
\cite{hubbard_honeycomb_0,hubbard_honeycomb_1,hubbard_honeycomb_2,hubbard_honeycomb_3},
presumably due to the importance
of long-range interactions and deviation from the nearest neighbor TB
approximation as $\rho$ increases.
On the other hand, according to our quantum Monte Carlo (QMC) calculations, MFH
trial wave functions based on LDA estimation of $U/t$ do not provide the most
suitable fixed-node energies and the nodal structure of the simplest
TB trial wave functions works best for the whole range of system
parameters regardless of the underlying electronic or magnetic state.
\begin{figure}[h]
  \centering
  \includegraphics[width=0.925\columnwidth]{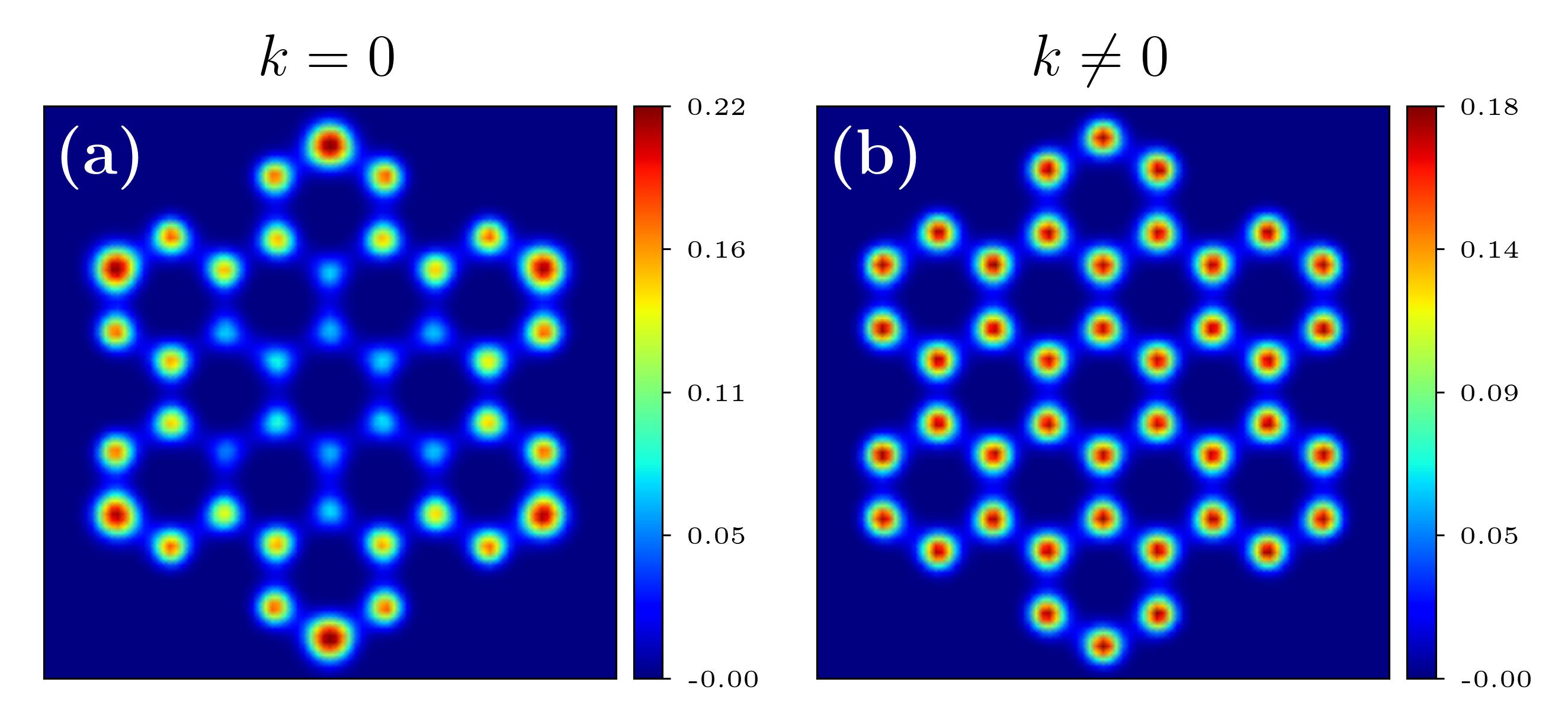}
  \caption{\label{effect_of_k} Effect of k on total electron
    densities for armchair hexagonal flake. $s = 1.4$, $\rho = 25$ nm and $N=42$
    using tight-binding trial wave function.
    (a) $k=0$.
    (b) $k = 3.56 \times 10^{-4}$ meV/nm$^2$.
  }
\end{figure}

Next, we focus on finite size effects. While $V_{0}$ is tuned to
imitate overall charge neutrality, long-range electron interactions
still affects the charge distribution inside the system, pushing the
electrons towards the edges and/or corners, as can be seen in
Fig.\ \ref{effect_of_k}a for the hexagonal armchair system with $N=42$,
$s = 1.4$ and $\rho = 25$ nm. While, in principle, $V_{0}$ can be decreased
further to achieve charge uniformity, this would localize the
electrons too strongly to their sites and make the system negatively
charged unlike in real graphene systems. An alternative
solution is to apply a quadratic gate potential controlled by the
parameter $k$ in Eq.\ (\ref{hamiltonian}).  When $k > 0$, quadratic gate
potential attracts the electrons towards the center of the system so
that the charge uniformity is preserved, and finite size effects are
minimized, as shown in Fig.\ \ref{effect_of_k}b
for $k = 3.56 \times 10^{-4}$ meV/nm$^2$.

\begin{figure}[h]
\centering
\includegraphics[width=0.975\columnwidth]{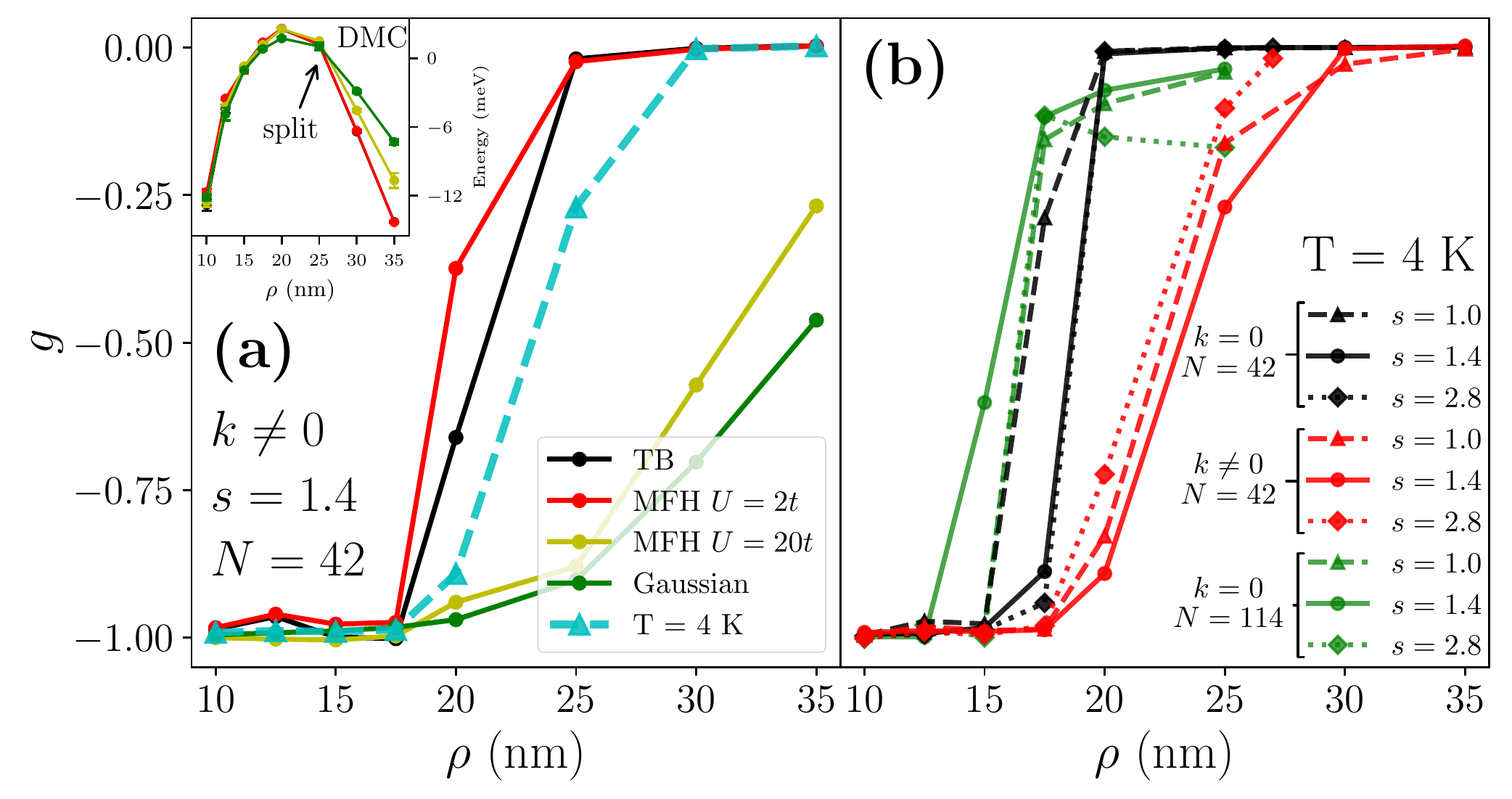}
\caption{\label{gate_dmc_results}
    Armchair hexagonal flake results.
    (a) Extrapolated spin-spin correlation function for $s = 1.4$, $k = 3.56 \times 10^{-4}$ meV/nm$^2$ and $N=42$.
    (a) inset is DMC energy results shifted by a constant value.
    (b) Extrapolated spin-spin correlation function, weighted averages at $T = 4$ K plotted for
    several $k$, $s$, and flake sizes.
}
\end{figure}
In order to understand more in detail the dynamics of the transition
from AF insulator to metallic phase including effects of charge
nonuniformity and quantum well potential sharpness, we consider a
real space spin-spin correlation function $g$ normalized by the
density-density correlations,
$g = \langle s_{i} s_{j} \rangle / \langle n_{i} n_{j} \rangle$
where $s_{i}$ and $n_{i}$ are the average total spin and total
electron densities on site i within a radius $r$, and $i, j$ are the
nearest neighbor sites. We used $r = a / 2$ in all spin-spin
correlation calculations, where $a$ is the lattice constant.  Output
values of the function remain in [-1,1] range, with $g = -1$
corresponding to AF and $g = 0$ corresponding to metallic
configuration ($g = 1$ means that all spins are in the same direction,
which does not happen in the subspace $S_z=0$ considered here). In
Fig.\ \ref{gate_dmc_results}a, $g$ is plotted against $\rho$ obtained
from different trial wave functions, for $s=1.4$ and
$k = 3.56 \times 10^{-4}$ meV/nm$^2$ to obtain charge uniformity. We
have also added a weighted average of all trial wave functions
using Boltzmann distribution at $T=4$ K, representing
the ground state, to ensure that no effects are
missed when various trial wave functions lead to same ground state
energies within statistical noise. The emerging picture is that the
system remains strongly AF between $\rho=10-18$ nm, then starts fading
smoothly, finally reaching fully metallic regime around $\rho=30$
nm. We note that the DMC energies for $k>0$ split around $\rho=26$ nm
(see the inset of Fig.\ \ref{gate_dmc_results}a), in contrast with the
$k=0$ results in Fig.\ \ref{first_result}b where the split occurs near
$\rho=18$ nm. Figure \ref{gate_dmc_results}b summarizes all 
weight averaged $g$ values obtained for different $k$ and $s$
values, and for both flake sizes.
Interestingly, for $k=0$ the transition from AF to metallic
regime is sharper than for charge uniform systems ($k>0$), albeit at
lower $\rho$ values.
Sharpness $s$ and system size $N$, on the other hand,
does not have a significant effect on the transitions.

\begin{figure}[h]
\centering
\includegraphics[width=0.90\columnwidth]{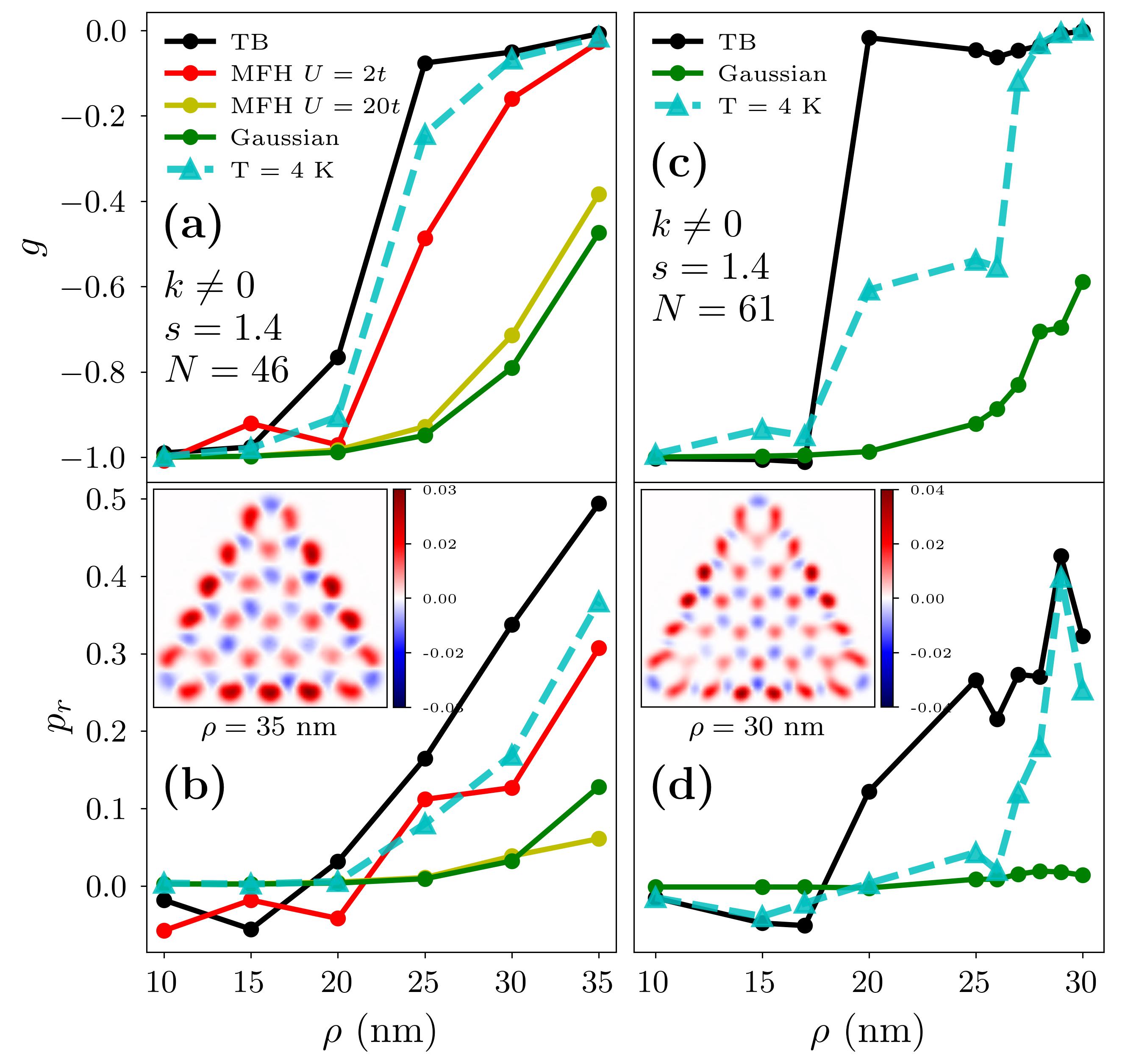}
\caption{\label{triangle}
    Triangular flake with zigzag edges results, $s=1.4$ and
    $k = 2.7 \times 10^{-4}$ meV/nm$^{2}$ plotted for several
    trial wave functions and different flake sizes.
    (a, c) Extrapolated spin-spin correlation function for 46 and 61 sites.
	(b, d) Extrapolated edge polarization function for 46 and 61 sites.
    (b inset) Extrapolated spin density results for $\rho = 35$ nm, $N=46$, using tight-binding trial wave function.
    (d inset) Extrapolated spin density results for $\rho = 30$ nm, $N=61$, using tight-binding trial wave function.
}
\end{figure}
For triangular zigzag structures, spin polarized edges even in the
metallic phase is expected due to the imbalance between the two
sublattices \cite{Rossier2007,Kaxiras2008,GQD_devrim_triangle,graphene_experimental_triangular_magnetism,AG_theoretical_yasser},
with a non-zero ground state total spin $S_z=(N_A-N_B)/2$ where $N_A$
and $N_B$ are number of $A$ and $B$ sublattice atoms, according to
Lieb's theorem \cite{Lieb}. 
In our AG flake, the total spin is $S_z=2$ for $N=46$ and $S_z=5/2$ for $N=61$.
The analysis of the spin-spin correlation functions
for $s=1.4$ and $k = 2.7 \times 10^{-4}$ meV/nm$^{2}$ leads to a similar
picture as before; the system with $N=46$ sites
is perfectly AF between $\rho=10-20$ nm,
then smoothly vanishes between $\rho=20-35$ nm, before becoming
completely metallic. 
For $N=61$ sites, although the transition is not as
smooth presumably due to larger statistical fluctuations at large system size,
a similar picture emerges (Fig.\ \ref{triangle}c).
In order to understand the edge
magnetization during the transition, we consider an edge polarization
ratio defined as
\begin{align}
p_r = \frac { \langle | s_{i\in edge} | \rangle - \langle | s_{i\in bulk} | \rangle }  { \langle | s_{i\in edge} | \rangle + \langle | s_{i\in bulk} | \rangle },
\end{align}
which gives one if only edge sites are polarized, zero if edge and
bulk sites are equally (un)polarized.  In Fig.\ \ref{triangle}b, edge
polarization ratio increases as $\rho$ increases, indicating that
spins are polarized more at the edges as the system goes into metallic
phase similar to real triangular zigzag graphene quantum dots, and
consistent with the spin density results in 
Fig.\ \ref{triangle}b and \ref{triangle}d insets, demonstrating
a metallic phase with edge-polarized spins at $\rho=35$ nm
for 46 sites and at $\rho=30$ nm for 61 sites. Additionally,
in Fig.\ \ref{triangle}, weighted average
analysis also confirms that the triangular
zigzag AG flake undergoes a metal-insulator transition between 
$\sim 18$ and $30$ nm dot radius. However, the edge polarization transition
occurs slower compared to the metal insulator transition.

In summary, we have shown that a metal - antiferromagnetic insulator transition
occurs in nanopatterned GaAs based artificial graphene structures
including up to $N=114$ electrons with
lattice constant $a=50$ nm as a function of site radius, using
accurate variational and diffusion Monte Carlo methods, within the
Dirac regime as confirmed by our density functional calculations. Our
approach where a simple tight-binding trial wave function combined
with a Jastrow factor is found to be sufficient to account for
electron correlation effects in both metallic and antiferromagnetic regimes, allows
direct modelling of system parameters, making it possible to
systematically investigate the effects of edge type, geometry, size, quantum
well shape and gate potentials. We have shown that the steepness and the
crossover $\rho$ value of the phase transition is affected by charge
nonuniformity due to finite size effects. For triangular structures
exhibiting magnetized edge states, the edge polarization transition is
shown to occur more slowly compared to the metal insulator transition.

\begin{acknowledgments}
  We thank C. J. Umrigar for his endless support for CHAMP \cite{Cha-PROG-XX} program
  with which our QMC simulations have been performed, Pawel Hawrylak and Yasser Saleem
  for valuable conversations. This work was supported by The Scientific and
  Technological Research Council of Turkey (TUBITAK) under the 1001 Grant
  Project Number 119F119.  The numerical calculations reported in this study were
  partially performed at TUBITAK ULAKBIM, High Performance and Grid Computing
  Center (TRUBA resources).
\end{acknowledgments}

\bibliography{main}

\end{document}